\documentclass[a4paper,12pt]{article}
\usepackage{amsmath}
\usepackage{pstricks}
\usepackage{pst-node}
\usepackage[ansinew]{inputenc}
\usepackage{amssymb,amsmath}
\usepackage{amsfonts}
\usepackage{epsfig}
\usepackage[english]{babel}
\usepackage{graphics}

\def\p{\partial}
\def\px{\partial_x}
\def\py{\partial_y}
\def\a{\alpha}
\def\b{\beta}
\def\g{\gamma}

\def\o{\omega}

\def\wt{\widetilde}

\def\vp{\varphi}

\font\Sets=msbm10

 \def\Rational {\hbox{\Sets Q}}

\def\be{\begin{equation}}       \def\ba{\begin{array}}

\def\ee{\end{equation}}         \def\ea{\end{array}}

\def\bea {\begin{eqnarray}}      \def\eea {\end{eqnarray}}

\def\bean{\begin{eqnarray*}}    \def\eean{\end{eqnarray*}}

\def\pa  {\partial}

\def\eps{\varepsilon}

\def\RA {\ \Rightarrow\ }

\def\<{\langle} \def\({\left(}  \def\>{\rangle} \def\){\right)}

\newtheorem{exi}{Example}

\author{E. Kartashova$^{\dag}$, O. Rudenko$^*$\\
$\dag$RISC, J. Kepler University, Altenbergerstr. 69, 4040 Linz, Austria\\
$*$ Dep. Chem. Phys.,  Weizmann Institute of
Science, Rehovot 76100, Israel \\
e-mails: lena@risc.uni-linz.ac.at,\\
oleksii@wisemail.weizmann.ac.il}

\title{Invariant Form of BK-factorization and its Applications}

\begin{document}
\date{}
\maketitle


\section{Introduction}

Factorization of ordinary and partial linear differential operators
(LODOs and LPDOs)
 is a very well-studied problem and a lot of  pure existence
theorems are known. For LODOs it is proven that a factorization is
unique while for LPDOs even uniqueness is not true any more and in
fact parametric families of factorizations can be constructed for a
given LPDO as will be demonstrated below. First {\bf constructive
method} of factoring second order hyperbolic LPDO in the form \be
\label{Lap}\p_x \p_y + a\p_x + b\p_y + c, \ee belong to Laplace who
formulated it in terms of invariants $\hat{a}= c- ab -a_x \quad
\mbox{and} \quad \hat{b}=c- ab  -b_y$ now called Laplace invariants.
An operator (\ref{Lap}) is factorizable  if at least one of its
Laplace invariants is equal to zero. Various  algorithms are known
now for factoring LPDOs over different differential fields beginning
with
the simplest field of rational functions \cite{beke94}.\\

Recently two papers (\cite{gs}, \cite{bk2005}) on factoring
arbitrary order  LPDOs have been published. In \cite{gs}  a
modification of well-known Hensel lifting algorithm (see, for
instance, \cite{Kalt}) is presented and {\bf sufficient conditions}
for the existence of intersection of principal ideals are given.
These results are applied then to re-formulate the factorization
formulae for  second and third order operators from the ring
$D=\Rational(x,y)[\px, \py]$ obtained by
 Miller (1932)  in terms of principal intersections. Authors do not
 claim to  construct general algorithm of factoring arbitrary
 order LPDOs but say that this approach is applied to
"execute a complete analysis of factoring and solving a second order
operator in two variables. Some results on factoring third-order
operators are exhibited". On the other hand, this approach can
probably be generalized to many variables.\\

 In \cite{bk2005} {\bf necessary and sufficient conditions} are given for factoring of
bivariate LPDOs of arbitrary order with coefficients being arbitrary
smooth functions. In \cite{k2006} it was shown that this procedure
called now BK-factorization principally can not be generalized on
the case of more than two variables. In was also shown that
conditions of factorization found  in \cite{bk2005} are invariants
under gauge transformation and classical Laplace invariants are
particular case of this generalized invariants. In this paper we
re-formulate BK-factorization in more suitable for applications
invariant form and illustrate it with a few examples, give a sample
of the symbolical implementation of this method in MATHEMATICA and
also discuss some possibilities to use this method for approximate
factorization of LPDOs.

\section{Helsel descent and BK-factorization}

We begin this paper with a brief comparison of Hensel descent and
BK-factorizations in order to show merits and draw-backs of each
method, just for complicity of presentation.

\begin{itemize}
\item{}In \cite{bk2005} factorization of a bivariate LPDO is looked for in
the form
\begin{equation}\label{An}
A_n=\sum_{j+k\le n}a_{jk}\partial_x^j\partial_y^k =(p_1\partial_x +
p_2 \partial_y + p_3)\Big(\sum_{j+k <
n}p_{jk}\partial_x^k\partial_y^j\Big)
\end{equation}
and in \cite{gs} - operator of $m \geq 2$ independent variables is
regarded and for $m=2$ factorization is looked for in the form
\begin{equation}
A_n=\sum_{j+k\le n}a_{jk}\partial_x^j\partial_y^k = \Big(\sum_{j+k
\leq l}p_{jk}\partial_x^k\partial_y^j\Big) \Big(\sum_{j+k \leq
r}p_{jk}\partial_x^k\partial_y^j\Big)  \ \ \mbox{with} \ \ l+r=n.
\end{equation}
\item{} In \cite{bk2005} coefficients $a_{jk}$ are arbitrary smooth
functions, for instance trigonometric functions; in \cite{gs}
conditions for reducibility of an operator are studied when
"coefficients are from a universal field of zero characteristic" ,
while "studying factorization algorithms we will assume that the
input operators are from the ring
$\Rational(x_1,...,x_m)[\p_1,...,\p_m]$." This suggestion is
necessary: "From now on the coefficients of a given second-order
operator are assumed to be from the base field $\Rational(x,y)$.
This is necessary if the goal is to obtain constructive answers
allowing to factorize large classes of operators" (\cite{gs},
Sec.3); "In this section we study third-order LPDOs from the ring
$D=\Rational(x,y)[\px, \py].$" (\cite{gs}, Sec.4).

\item{} In \cite{bk2005} it was shown that in generic case factorization
can be constructed explicitly and algebraically, while in \cite{gs}
(Sec.5) it is concluded that "the factorization problem for second-
and third-order differential operators in two variables has been
shown to require the solution of a partial Riccati equation, which
in turn requires to solve a general first-order ODE and possibly
ordinary Riccati equation. The bottleneck for designing a
factorization algorithm for a LPDO is general first-order ODE which
make the full problem intractable at present because in general
there are no solution algorithm available."

\item{} In \cite{k2006} it is pointed out that BK-factorization
procedure has to be modified in some way (presently unknown to
authors) in order to proceed with operators of more than 2
independent variables, while in \cite{gs} (Sec.5) it is written that
"some of the results described in this article may be generalized to
any number of independent variables."

\item{} In \cite{k2006} it shown that BK-factorization procedure gives
rise to construction of the whole class of generalized invariants
particular case of them  being classical Laplace invariants. This
leads to a possibility to factorize simultaneously the whole class
of operators equivalent under gauge transformation (see next
Section) while Helsel descent is used for factoring of a one
specific operator.

\end{itemize}

We summarize all this in the Table below.\\

\begin{tabular}{|c||c||c|}
  \hline 
 Property / Method & BK-factorization & Hensel descent \\
\hline
Order of operator & $n$\ &  $n$\\
\hline
Operator's coefficients & arbitrary smooth functions & rational functions\\
\hline
Number of variables & 2 &  possibly $>2$\\
\hline
Conditions of factorization & necessary and sufficient  & sufficient\\
\hline Form of factorization & factors of order 1 and $(n-1)$  & factors of order $k$ and $(n-k)$ \\
\hline Formulation in terms of & explicit formulae for fact. coefficients  & existence of ideals intersection \\
\hline
\end{tabular}

In the next Sections we demonstrate some other interesting
properties of BK-factorization - first of all, that it has invariant
form and can be used therefore to factorize simultaneously the whole
classes of equivalent LPDOs. Second, the use of this invariant form
of BK-factorization for construction of approximate factorization
for LPDEs to be solved numerically.

\section{Invariant Formulation}

We present here briefly main ideas presented in \cite{bk2005},
\cite{k2006} beginning with the definition of equivalent operators.

\paragraph{Definition} The operators $A$, $\tilde A$ are called
equivalent if there is a gauge transformation that takes one to the
other:
$$
\tilde Ag= e^{-\vp}A(e^{\vp}g).
$$
BK-factorization is then pure algebraic procedure which allows to to
construct explicitly a factorization of an arbitrary order LPDO $A$
in the form
$$
A:=\sum_{j+k\le n}a_{jk}\partial_x^j\partial_y^k=L \circ
\sum_{j+k\le (n-1)}p_{jk}\partial_x^j\partial_y^k
$$
with first-order operator $L=\pa_x-\o\pa_y+p$ where $\o$ is {\bf an
arbitrary  simple root} of the characteristic polynomial \be
\label{char} \mathcal{P}(t)=\sum^n_{k=0}a_{n-k,k}t^{n-k}, \quad
\mathcal{P}(\o)=0. \ee 
Factorization is possible then for each simple root $\tilde{\o}$ of
(\ref{char}) {\bf iff}

for $n=2 \ \ \RA l_2=0,$

for $n=3 \ \ \RA l_3=0, \ \& \ l_{31}=0,$

for $n=4 \ \ \RA l_4=0, \ \& \ l_{41}=0, \ \& \ l_{42}=0,$

and so on. All functions $l_2, \ \ l_3, \ \ l_{31}, \ \  l_4 \ \
l_{41}, \ \ l_{42}, ...$ are explicit functions of $a_{ij}$ and
$\tilde{\o}$.

\paragraph{Theorem}  All $l_2, l_3, l_{31}, ....$ are {\bf invariants}
under gauge transformations.

\paragraph{Definition} Invariants $l_2, l_3, l_{31}, ....$ are
called generalized invariants of a bivariate operator of arbitrary
order.\\

In particular case of the operator (\ref{Lap}) its generalized
invariants coincide with Laplace invariants.

\paragraph{Corollary} If an operator $A$ is factorizable, then all
operators  equivalent to it, are also factorizable.\\

 As the first step of BK-factorization,
coefficients  $p_{ij}$ are computed as solutions of some
 system of algebraic equations. At the second step, equality to zero of all generalized invariants
 $l_{ij}=0$ has to be checked so that
{\bf no differential equations are to be solved} in generic case.
Generic case corresponds to a simple root of characteristic
polynomial, and each simple root generates  corresponding
factorization.  Moreover, putting  some restrictions on the
coefficients of the initial LPDO $a_{i,j}$ as functions of $x$ and
$y$, one can describe {\it all} factorizable operators in a given
class of functions (see Example 5.3 in \cite{bk2005}). The same
keeps true for all operators equivalent to a given one. Equivalent
operators are easy to compute:
$$e^{-\vp} \px e^{\vp}= \px+\vp_x, \quad e^{-\vp} \py e^{\vp}=
\py+\vp_y,$$ $$e^{-\vp} \px \py e^{\vp}= e^{-\vp} \px e^{\vp}
e^{-\vp} \py e^{\vp}=(\px+\vp_x) \circ (\py+\vp_y)$$ and so on. Some
examples:
\begin{itemize}
 \item{}$A_1=\px \py + x\px + 1= \px(\py+x), \quad
l_2(A_1)=1-1-0=0;$

 \item{}$A_2=\px \py + x\px + \py +x + 1, \quad
A_2=e^{-x}A_1e^{x};\quad l_2(A_2)=(x+1)-1-x=0;$

 \item{}$A_3=\px \py + 2x\px + (y+1)\py +2(xy +x+1), \quad
A_3=e^{-xy}A_2e^{xy}; \quad l_2(A_3)=2(x+1+xy)-2-2x(y+1)=0;$

\item{} $A_4=\px \py +x\px + (\cos x +1) \py + x \cos x +x +1, \quad
A_4=e^{-\sin x}A_2e^{\sin x}; \quad l_2(A_4)=0.$
\end{itemize}

Generic case which can be treated pure algebraically by
BK-factorization corresponds to {\bf a simple root of characteristic
polynomial}. Each multiple root leads to necessity of solving some
Ricatti equation(s) (RE). If appeared RE happens to be solvable,
such a root generates a parametric family of factorizations for a
given operator. For instance, well-known Landau operator
$$
\p_{xxx}^3+x\p_{xxy}^3+2\p_{xx}+(2x+2)\p_{xy}^2+\p_x+(2+x)\p_y
$$
has characteristic polynomial  with one distinct root $\o_1=-x$ and
one double root $\o_{2,3}=0.$ Factorization then has form
$$
(\p_x+r)(\p_x-r+2)(\p_x+x\p_y)
$$
where $r$ is a solution of Ricatti equation $$ 1-2r+\p_x(r)+r^2=0$$
which is easily solvable: $$r =1+\frac{1}{x+Y(y)}$$ with arbitrary
smooth function $Y(y)$ of one variable $y$ so that factorization has
form
$$
A=(\p_x+1+\frac{1}{x+Y(y)})(\p_x+1-\frac{1}{x+Y(y)})(\p_x+x\p_y).
$$

Notice that to factorize  {\bf an ordinary differential operator} it
is always necessary to solve some RE. Nevertheless, just formal
application of BK-factorization will produce all the linear factors
in the case when corresponding RE are solvable. For instance, the
factorization has been constructed  in \cite{usa}
$$
  x \partial_{xxx} + (x^2-1)\partial_{x x} - x \partial_{x}
  +  \frac{2}{x^2} - 1 =
  (\partial_{x}+ \frac{x^2-1}{x})(x \partial_{x}-\sqrt2)(\partial_{x} + \frac{\sqrt2
  -1}{x}).
$$
while both RE appearing at the intermediate steps are solvable.\\

These two last examples show the main difference between factorizing
of ordinary and partial differential operators - LODO has always
unique factorization while LPDO may has many. An interesting
question here would be to compute the exact number of all possible
factorizations of a given LPDO into all linear factors (its upper
bound is, of course,  trivial: $\ \ n!$).  A really challenging
 task in this context would be to describe some additional
conditions on the coefficients of an initial operator
 which lead to solvable RE.

\section{Left and Right Factors}

Factorization of an operator is the first step on the way of solving
corresponding equation. But for solution we need {\bf right} factors
and BK-factorization constructs {\bf left} factors which are easy to
construct. On the other hand, the existence of a certain right
factor of a LPDO is
  equivalent to the existence of a corresponding left factor of
  the {\it transpose\/} of that operator.
 Moreover taking
  transposes is trivial algebraically, so there is also nothing
  lost from the point of view of algorithmic computation.

\paragraph{Definition}
The transpose $A^t$ of an operator
$
A=\sum a_{\a}\p^{\a},\qquad \p^{\a}=\p_1^{\a_1}\cdots\p_n^{\a_n}.
$
is defined as
$$
A^t u = \sum (-1)^{|\a|}\p^\a(a_\a u).
$$
and the identity
$$
\p^\g(uv)=\sum \binom\g\a \p^\a u\,\p^{\g-\a}v
$$
implies that
$$
A^t=\sum (-1)^{|\a+\b|}\binom{\a+\b}\a (\p^\b a_{\a+\b})\p^\a.
$$

Now the coefficients are
$$
A^t=\sum \wt a_\a \p^\a,$$
$$ \wt a_\a=\sum (-1)^{|\a+\b|}
\binom{\a+\b}{\a}\p^\b(a_{\a+\b}).
$$
with a standard convention for binomial coefficients in several
variables, e.g. in two variables
$$
\binom\a\b=\binom{(\a_1,\a_2)}{(\b_1,\b_2)}=\binom{\a_1}{\b_1}\,\binom{\a_2}{\b_2}.
$$
In particular, for order 2 in two variables the coefficients are
$$
\wt{a}_{jk}=a_{jk},\quad j+k=2;
 \wt{a}_{10}=-a_{10}+2\px a_{20}+\py
a_{11}, \wt{a}_{01}=-a_{01}+\px a_{11}+2\py a_{02},$$$$
\wt{a}_{00}=a_{00}-\px a_{10}-\py a_{01}+\px^2 a_{20}+\px\px
a_{11}+\py^2 a_{02}.$$

For instance,  the operator \be
\label{fact}\p_{xx}-\p_{yy}+y\p_x+x\p_y+\frac{1}{4}(y^2-x^2)-1 \ee
is factorizable as
$$\big[\px+\py+\tfrac12(y-x)\big]\,\big[...\big]$$
and its transpose $A_1^t$ is factorizable then as
$$\big[...\big]\,\big[\px-\py+\tfrac12(y+x)\big].$$

Implementation of the BK-factorization for bivariate operators of
order $n \le 4$ is therefore quite straightforward and has been done
in MATHEMATICA while all roots of characteristic polynomial are
known in radicals. For instance, for the operator (\ref{fact}) with
2 simple roots we get one factorization
$$
\big[\px-\py+\tfrac12(y+x)\big]\,\big[\px+\py+\tfrac12(y-x)\big].
$$
corresponding to the first root while in the case of the second
root,
generalized invariant is equal to 2.\\

If $n \ge 5$ the problem is generally  not solvable in radicals and
very simple example of non-solvable case is: $x^5 - 4x - 2=0$. Thus,
to find solutions in radical for $n>4$ one needs some constructive
procedure of finding solvable Galois group but this lies beyond the
scope of the present paper.

\section{Approximate Factorization}
An interesting possible application of the invariant form of
BK-factorization is to use it for construction of approximate
factorization of a given LPDE, in the case when exact factorization
of corresponding LPDO does not exists. Indeed, as a results of
BK-factorization one gets

(1) {\bf factorization coefficients} $\{p^{(i)}_{ij}\} $ for the
$i$-th factorization of a given operator, and

(2) {\bf generalized invariants} $ l^{(i)}_2p^{(i)}_{ij}, $ with all
$p^{(i)}_{ij}, l^{(i)}_{(k_j)}$ being explicit functions of the
coefficients of initial operator
 $a_{ij}.$\\

In numerical simulations coefficients  $a_{ij}$ of the equation  are
always given with some non-zero accuracy, say $\varepsilon
>0,$ which means that it is enough to construct an approximate
factorization in the following sense.  One has to find restrictions
on the coefficients $a_{ij}$ of an initial LPDO which provide
$|l^{(j)}_{k_j}| < \eps$  with a given accuracy $0<\eps<<1$. Many
different strategies are possible here, we just give a brief sketch
of two approaches we are working on right now:

\subsection{Quantifier Elimination}

We illustrate this idea on the simple example of a
 hyperbolic operator $\p_{xx}-\p_{yy}+a_{10}\px+a_{01}\py+a_{00}$  with linear polynomial
coefficients. {\bf What we have is:}

$$
a_{00}(x,y)=b_3x+b_2y+b_1,$$$$ a_{10}(x,y)=c_3x+c_2y+c_1,$$$$
a_{01}(x,y)=d_3x+d_2y+d_1; $$
 a function constructed from general
invariants $$ \mathcal{R}=
 \frac{ s_3-s_2}{2}+ \frac{( s_3x+s_2y+s_1)^2}{4}$$ with
 $s_i=c_i-d_i$.

{\bf What we need is:}

To find some function(s) $F=F(a_{ij})$ such that if $F(a_{i,j})=0,$
then $$ -\varepsilon<a_{00} - \mathcal{R}<\varepsilon, \ \ \mbox{for
some constant} \ \ 0<\varepsilon<<1,
$$
i.e. to find some conditions on the initial polynomials which
provide that function $\mathcal{R}$ differs not too much from one
these polynomials, namely $a_{00}$.\\

 Notice that simple symmetry
considerations
 allowed us to reduce
number of variables needed for CAD calculations. Initially we had 9
variables $b_3, b_2, b_1, c_3, c_2, c_1, d_3, d_2, d_1 $ but in fact
it is enough to regard only 6 variables $s_1, s_2, s_3, b_1, b_2,
b_3$. Nevertheless, the computation time may become crucial while
using this approach due to the substantial number of variables. On
the other hand, this approach allows us work generally on the
operator level including initial and/or boundary conditions first at
some later stage.

\subsection{Auxiliary Operator}

Another approach is to construct a new auxiliary operator with
coefficients $\wt{a}_{ij}= f(x,y))a_{ij}$ for all or for some of the
coefficients $a_{ij}$ of the initial operator, keep invariants
(almost) equal to zero and find function(s) $f(x,y)$ minimizing the
differences between the coefficients of initial and new operators.
In this way an auxiliary operator is constructed which can be
regarded as an approximate operator for the initial operator. Of
course, it does not mean that solutions of the initial and
approximate operators will be also close  but simple properties of
linear operators show that it is necessary (but not sufficient!)
step on the way of construction of a good approximate solution of a
given LPDE - in the case of a well-posed problem, of course. In
particularly, it means that one have to introduce proper metrics in
the space of operators {\bf and} in the space of solutions. Choice
of the both metrics and of a function $f$ will depend on (1)
coefficients of the initial operator; (2) class of functions in
which we are looking for a solution; (3) initial and/or boundary
conditions.\\

To demonstrate all this  let us regard two different un-factorizable
modifications of the  operator (\ref{fact}):
\be
\label{fact1}A=\p_{xx}-\p_{yy}+y\p_x+x\p_y+\frac{1}{2}(y^2-x^2)-1
\ee
with $l_2(A)=\frac{1}{4}(y^2-x^2)$
 and
  \be
\label{fact2}B=\p_{xx}-\p_{yy}+\sin{y} \ \p_x+\cos{x} \
\p_y+\frac{1}{2}(\sin^2{y}-\cos^2{x}) \ee 
with $l_2(B)=\frac {1}{2} (\cos y - \sin x)$ (see Fig.1). One can
see immediately that $l_2(B)$ is a bounded function of two variables
and  $l_2(A)$ is an unbounded. This means that quite different
choice of function $f$ is needed for these two cases in order to
minimize the invariants. Influence of initial/boundary conditions is
now also very clear - for instance, best approximation of $l_2(B)$
can be obtained in the narrow belts of the lines parallel to one of
the coordinate axis while for $l_2(A)$ these directions are in no
way special.\\

 \begin{figure}
\begin{center}
\includegraphics[width=6.5cm,height=5cm]{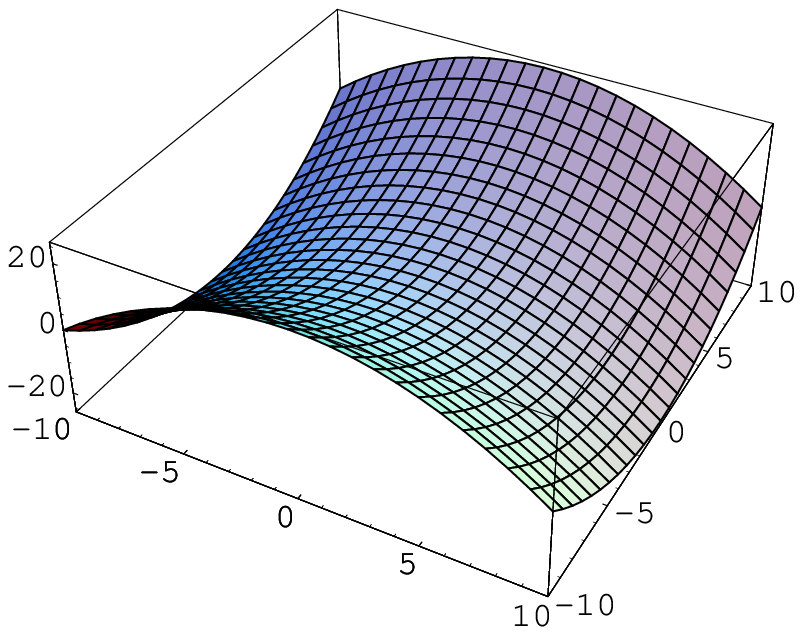}
\includegraphics[width=6.5cm,height=5cm]{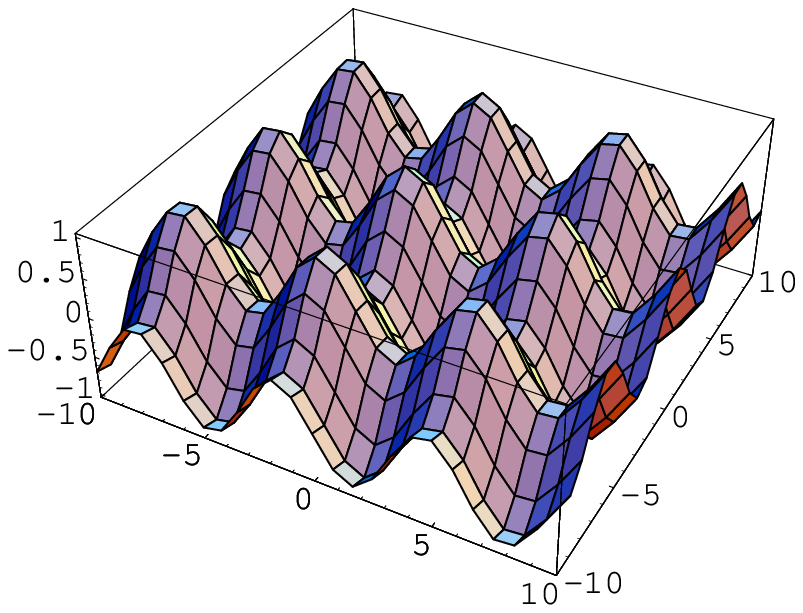}
\end{center}
\caption{\label{f:4} Invariant $l_2(A)=\frac{1}{4}(y^2-x^2)$ (left)
and invariant $l_2(B)=\frac {1}{2} (\cos y - \sin x)$  (right), in
the domain $-10 \le x,y \le 10$}
\end{figure}

 \begin{figure}
\begin{center}
\includegraphics[width=6.5cm,height=5cm]{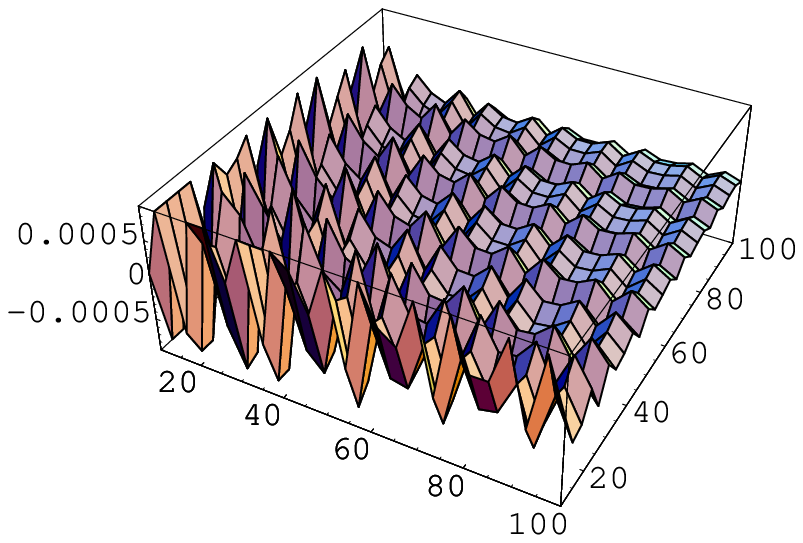}
\includegraphics[width=6.5cm,height=5cm]{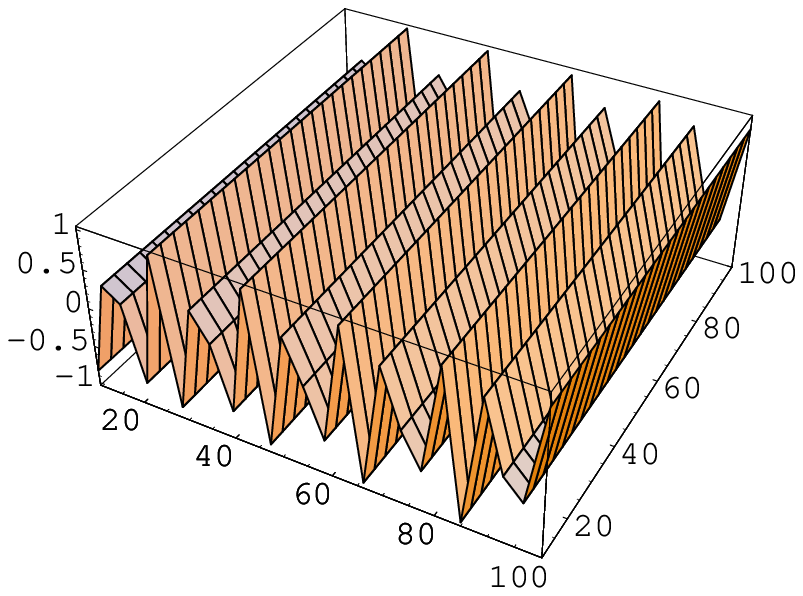}
\includegraphics[width=6.5cm,height=5cm]{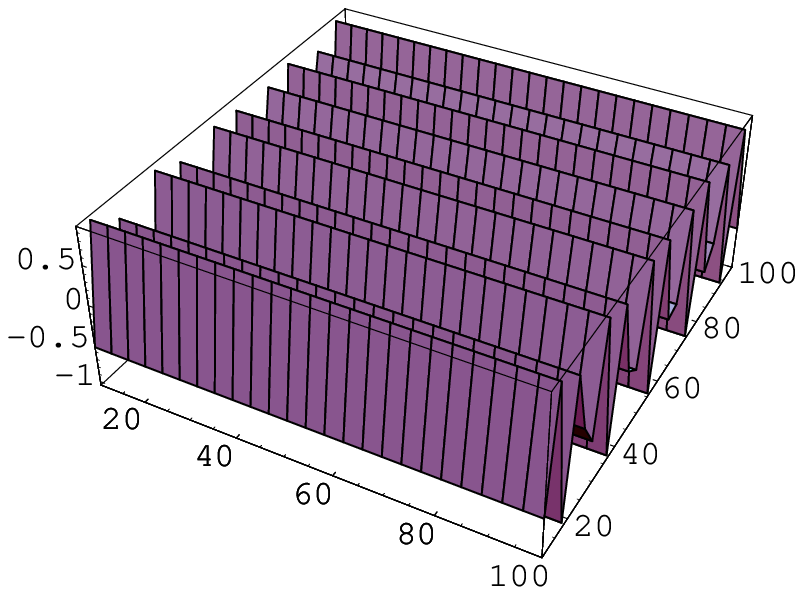}
\includegraphics[width=6.5cm,height=5cm]{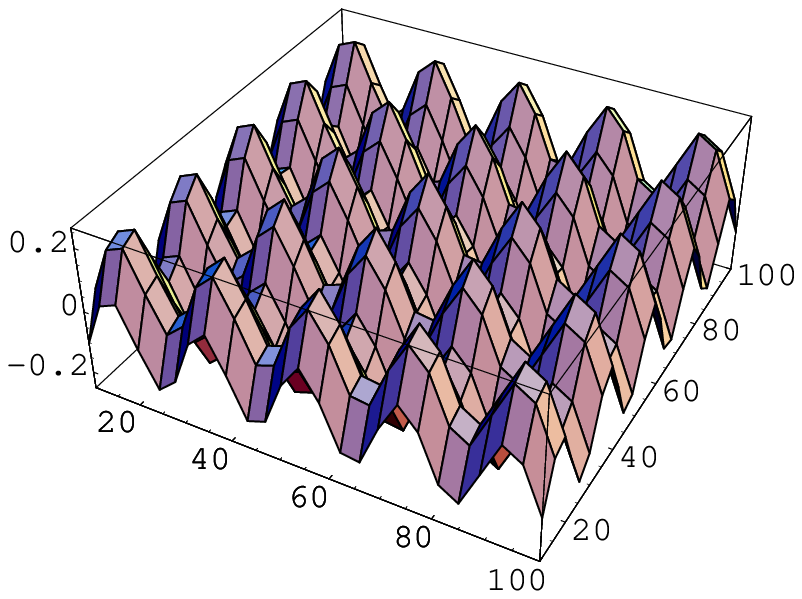}
\end{center}
\caption{\label{f:4} Upper panel:  $l_2(\tilde{B})$ (left) and
$a_{10}-\tilde{a}_{10}$ (right); lower panel:
$a_{01}-\tilde{a}_{01}$ (left) and $a_{00}-\tilde{a}_{00}$ (right);
in the domain $10 \le x,y \le 100$}
\end{figure}

To construct a sample  of such an approximate factorization for the
operator (\ref{fact2}) we just suppose intuitively that auxiliary
operator $\tilde{B}$ is "good" if its coefficients differ from the
coefficients of (\ref{fact2}) not much,
 and its invariant is small. Our MATHEMATICA implementation of the
 BK-factorization includes  simple graphic functions to display the
 differences between all the parameters of the initial and auxiliary
 operators.  A choice of the function $f(x,y)=
\sin{\frac{1}{xy}}$ gives an auxiliary operator $\tilde{B}$ of the
form
  \be
\label{fact3}\tilde{B}=\p_{xx}-\p_{yy}+\sin{y} \sin{\frac{1}{xy}} \
\p_x+\cos{x} \sin{\frac{1}{xy}} \
\p_y+\frac{1}{2}(\sin^2{y}-\cos^2{x}) \sin{\frac{1}{xy}}. \ee 

It is demonstrated at the Fig.2  that for $\ \ 10 \le x,y \le 100 \
\ $ operator $\tilde{B}$ gives good enough approximation and
correspondingly approximate factorization of the initial operator
$B$ has form
$$
B \sim \Big [ \frac{1}{2}\Big( -\cos x \sin \frac{1}{xy} +  \sin
\frac{1}{xy} \sin y\Big) +\px + \py \Big] \Big [ \frac{1}{2}\Big(
\cos x \sin \frac{1}{xy} +  \sin \frac{1}{xy} \sin y\Big) +\px - \py
\Big]
$$
with $|l_2(\tilde{B})| \sim 5\cdot 10^{-4} .$ On the other hand, in
the domain $\ \ 0.001 \le x,y \le 1 \ \ $ qualitatively different
approximation is needed while in this domain $|l_2(\tilde{B})| \sim
10^2$ (see Fig. 3).

 \begin{figure}
\begin{center}
\includegraphics[width=6.5cm,height=5cm]{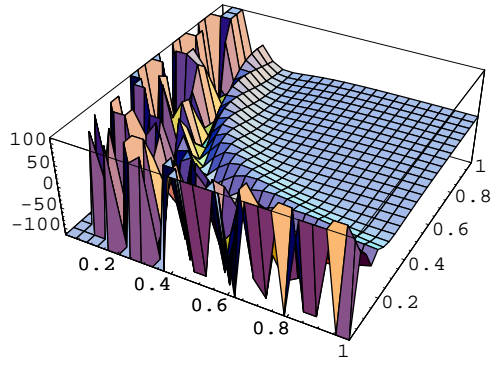}

\end{center}
\caption{\label{f:4} Invariant $l_2(\tilde{B})$ in the domain
$0.0001\le x,y \le 1$ }
\end{figure}

\section{Brief Discussion}

We presented here an invariant formulation of BK-factorization which
allows to factorize simultaneously  classes operators equivalent to
the initial one under gauge transformations. We also showed the
possibility to use the same procedure for the construction of the
approximate factorization of LPDE in the case when corresponding
LPDO  {\bf is not factorizable}. Obviously, if we get enough
approximate factorizations of the given LPDE with different solvable
first-order factors we can write out explicitly general solution of
the initial LPDE. Otherwise, one gets a chain of the linear
first-order equations
$$ A_{{i0},n}\psi_0=0, \ A_{{i0},n-1}\psi_1=\psi_0, \ .... $$
to be solved numerically which is a great numerical simplification,
of course, specially for higher order LPDEs. On the other hand,
while performing numerical simulations, one has to take into account
a lot of other factors, first of all,  initial and boundary
conditions. It would be a nontrivial task to include them into the
exact formulae given by BK-factorization. In order
 to estimate usefulness of this approach
from numerical point of view  we still have to answer all the
questions concerning computation time, stability, computation error,
etc. For instance, coming back to the example of approximate
factorization given in the previous section, one have to estimate
what is numerically more reasonable for a given set on initial and
boundary conditions - to solve numerically the system of equations
$$
\begin{cases}
[ \frac{1}{2}( \cos x \sin \frac{1}{xy} +  \sin
\frac{1}{xy} \sin y) +\px - \py ]\circ \psi_0=0 \\
[ \frac{1}{2}( -\cos x \sin \frac{1}{xy} +  \sin \frac{1}{xy} \sin
y) +\px + \py ]\circ \psi_1=\psi_0
\end{cases}
$$
or one equation $B\circ \psi=0.$ Some answers can be given by the
method presented in \cite{vl} where a symbolic approach is used to
generate automatically finite difference schemes for LPDEs and to
check their von Neumann stability. Some preliminary steps to be
taken in this direction might be following: (1) to take a
non-factorizable but solvable operator, for instance, $A_1=\px \py +
x\px + 2$, then LPDE $A_1(\psi)=0,$ has general solution
$$
\psi= -\p_x \Big( X(x)e^{-xy}+ \int e^{x(y'-y)}Y(y')dy'\Big)$$ with
two arbitrary functions $X(x)$ and $Y(y)$;
 (2) to
construct its approximate factorization $\wt{A}_1=L_1 \circ L_2;$
(3) to get computational schemes using \cite{vl} - for $A_1$ and
$\wt{A}_1$; (4) compute both numerically; (5) to compare results for
$A_1$ and $\wt{A}_1$ with the general solution for some classes of
initial data and for a fixed choice of computational scheme.

\section*{ACKNOWLEDGMENTS}

E.K. acknowledges the support of the Austrian Science Foundation
(FWF) under projects SFB F013/F1304 and Prof. Langer and Prof. Engl
for valuable discussions. O.R. acknowledges the support of the
US-Israel Binational Science Foundation.

\end{document}